\documentstyle[twocolumn,psfig,aps]{revtex}
\begin{document}
\draft    

\title{Quantum effects on dynamics of instabilities in Bose-Einstein
condensates}

\author{V. A. Yurovsky}

\address{ITAMP, Harvard-Smithsonian Center for Astrophysics, 60 Garden
Street, Cambridge MA 02138}

\address{Atomic Physics Division, Stop 8423, National Institute of
Standards and Technology, Gaithersburg, MD 20889}

\address{School of Chemistry, Tel Aviv University, 69978 Tel Aviv,
Israel}

\date{September 25, 2001}
\maketitle    

\begin{abstract} Dynamics of fluctuations in unstable
Bose-Einstein condensates is analyzed by the solution of
approximate operator equations. In the case of a condensate
with a negative scattering length the present treatment
describes a delay of collapse, in agreement with recent
experiments. In the case of a collision of two condensate
wavepackets it is shown that quantum effects lead to a Bose
enhancement of  elastic-scattering losses. In both cases the
noncondensate  atoms are formed as entangled pairs in
squeezed states.

\end{abstract}
\narrowtext

\section{Introduction}

Recent progress in the physics of Bose-Einstein condensation
(BEC) of atomic gases (see Refs.\ \cite{PW98,DGPS99} and references
therein) stimulated a further development of theoretical methods used
previously for the description of superfluidity, superconductivity,
and collective effects in heavy atoms and nuclei (see Refs.\
\cite{LP80,FV71,BB94,BR86}). These methods treat condensates as mean
fields, while interactions between particles, according to the
Bogolubov theory, lead to formation of fluctuations (see Refs.\
\cite{PB96,PBS98,G96,WWCH99}). In the case of a positive  scattering
length, the
condensate (mean field) is stable, while fluctuations can be treated
by using a Bogolubov transformation to  quasiparticle states.

Recent experiments \cite{SGWH99,GSPH00,R01,D01} drew attention
to the physics of condensates with a negative scattering length. In
this case, a free condensate (with a homogeneous mean field) is
unstable,  collapsing within a finite time \cite{DGPS99}, while a
trapped condensate  is stable at the low occupations of the experiment
\cite{SGWH99,GSPH00}.   In the JILA experiments on Rb \cite{R01,D01},
the effect of Feshbach  resonance (see Ref.\ \cite{TTHK99}) was used
for changing the sign of the  scattering length. This effect allows
the formation of a condensate with  a positive scattering length,
followed by a study of the dynamics of  collapse after applying a
sudden change of sign of the scattering length.  A formal extension of
the Bogolubov transformation to the case of   negative scattering
lengths would lead to complex energies of the  quasiparticles, but
actually such a transformation does not exist in this  case, as was
proven by Bogolubov (see Ref.\ \cite{BB94}).  Nevertheless, the
development of fluctuations can substantially affect  the dynamics of
an unstable condensate (see the theoretical analysis  \cite{DS01} of
the JILA experiments \cite{D01}).

The present paper describes the dynamics of fluctuations by using
the parametric approximation employed in Ref.\ \cite{VYA01} for a
model problem of a two-mode atom-molecule condensate, and in Ref.\
\cite{YBJ01} for a description of quantum effects on curve crossing.
This method is similar to the parametric approximation used in quantum
optics (see Ref.\ \cite{SZ97}), and in the description of interaction
of  electromagnetic and matter waves in Refs.\ \cite{Moore}. It is
also  closely related to the linearization approximation in quantum
soliton  theory (see Ref.\ \cite{HY00} and references therein) and
to the theory of quantum integrals of motion (see Refs.\ \cite{DM89}
and references therein). The
fluctuations  are treated here by a linearization of the exact quantum
equations of  motion for the field operators. In the case of a stable
BEC, the  resulting linear equations describe the motion of quantum
oscillators,  corresponding to the Bogolubov quasiparticles.  However,
in the case of unstable condensates, these equations describe
inverted oscillators, leading to a rapid growth of the fluctuations.
This growth remains super-exponential as long as the fluctuations do
not reach a macroscopic occupation. This dynamics offers a more
precise  picture compared to the exponential growth obtained from a
Liapunov  analysis of the classical mean field. Initially, the
dynamics can be  described as a quantum process of spontaneous decay
of the condensate into  the fluctuation vacuum (see Refs.\
\cite{VYA01,YBJ01}), accompanied by  the formation of entangled pairs
(see also Refs.\ \cite{VA01,AV01,PM01}). As in the case of
fluctuations in a stable condensate (see Ref.\ \cite{RGB01}), the
pairs are formed in two-mode squeezed states.

The formally calculated energy of a Bogolubov quasiparticle
becomes complex also when the energy of the excitation is lower than
the energy of a particle in the condensate state. This case can take
place if the condensate itself occupies an excited translational
state, as realized recently in the NIST experiments on four-wave
mixing of atomic  condensate waves \cite{D99}. The loss of condensate
atoms in such a situation was studied in  Ref.\ \cite{BTBJ00}, using
the method of complex scattering lengths,  without taking into account
second-quantized properties of the atomic  scattering. The results of
the present theory, which take into account these properties, are in
agreement with Ref.\ \cite{BTBJ00} at short collision times [see Eq.\
(\ref{ndotstat}) below]. However, at long collision times the  effects
of Bose enhancement lead to a substantial increase of the losses  [see
Eq.\ (\ref{ndotenh}) below]. Notice that the necessity of accounting
for the quantum effects was mentioned in Ref.\ \cite{BTBJ00} itself.
These effects should appear under the conditions of the NIST
experiments  \cite{D99}. The present method can describe only
qualitatively this case  of large losses, comparable in magnitude to
the initial occupation.  This case may be treated by the incorporation
of quantum effects  into the slowly-varying-envelope approximation
developed in  Ref.\ \cite{BTBJ00,TBJ00}.

Quantum effects on the loss incurred by collisions of two
condensate  wavepackets were studied in Ref.\ \cite{BTR01}, using an
approach  similar to the present one. However, the results of the
present work  disagree with those of Ref.\ \cite{BTR01}. The losses
quoted in  Ref.\ \cite{BTR01}, which have been evaluated only for
short collision  times, differ also from the results of Ref.\
\cite{BTBJ00}.  These disagreements are apparently due to some
unjustified neglections  in Ref.\ \cite{BTR01} (see discussion in
Sec.\ \ref{Gauss}).

As in the case of fluctuations in a ground-state condensate, the
scattered particles are formed here in entangled (squeezed) states.
Therefore the processes considered here may open up possibilities for
the formation of an entangled gas remained after the decay of a
condensate, or after the collision of wavepackets. The entangled gas
can have many application (see Ref.\ \cite{RGB01}).

This paper is organized as follows. Section \ref{Ground}
considers the dynamics of instabilities in a ground-state condensate
with a negative scattering length. The equations of motion for the
condensate and the fluctuations are derived in Sec.\ \ref{CondSep} by
using a projection onto a coherent state. The equations for the
fluctuations are solved in Sec.\ \ref{StabUnstMod}. In the  dependence
on the excitation energy, the fluctuations are divided into two
classes, of stable and unstable modes,  which are respectively
associated with regular and inverted oscillators  in Sec.\
\ref{RegInvOsc}.   The dynamics and statistical properties of the
instabilities are  considered in Sec.\ \ref{InstDyn}, and Sec.\
\ref{CondDepl} describes  the dynamics of condensate depletion.

The collision of two matter waves is studied in Sec.\
\ref{twocond}.  General equations for the condensate and the
fluctuations are derived in Sec.\ \ref{TCSep}. Section \ref{Loss}
considers elastic losses in collisions of two matter waves. The
collisions of  rectangular and Gaussian wavepackets are treated in
detail in Secs.\ \ref{Rect} and \ref{Gauss}, respectively.

A system of units in which Planck's constant is $\hbar =1$ is used
below.

\section{Ground-state condensate} \label{Ground}

\subsection{Separation the condensate and fluctuations}
\label{CondSep}

Consider a homogeneous  gas of bosons of mass $m$, in a finite quantization
box of  volume ${\cal V}$,
described by the following Hamiltonian in the momentum representation
(see Refs.\ \cite{PW98,DGPS99}),
\begin{eqnarray}
\hat{H}=&&\sum\limits^{}_{{\bf p}}\hat{\psi }^{\dag }\left( {\bf
 p},t\right) \Bigl\lbrack \epsilon \left( {\bf p}\right) \hat{\psi
 }\left( {\bf p},t\right)  \nonumber
\\
&&+{1\over 2{\cal V}}\sum\limits^{}_{{\bf q},{\bf p}^\prime }U\left(
 q\right) \hat{\psi }^{\dag }\left( {\bf p}^\prime ,t\right)
 \hat{\psi }\left( {\bf p}-{\bf q},t\right) \hat{\psi }\left( {\bf
 p}^\prime +{\bf q},t\right) \Bigr\rbrack  . \label{H}
\end{eqnarray}
Here $\hat{\psi }\left( {\bf p},t\right) $ is the annihilation
 operator for bosons in the momentum ${\bf p}$-mode,
$\epsilon \left( {\bf p}\right) ={\bf p}^{2}/\left( 2m\right) $
is the kinetic energy, and $U\left( q\right) $
describes the elastic scattering. In the model case of a zero-range
contact potential, $U$
is a constant, and can be expressed in terms of the elastic scattering
length $a_{a}$ as
\begin{equation}
U=4\pi a_{a}/m . \label{Ua}
\end{equation}
(A modification of the contact-potential model was recently proposed in
Ref.\ \cite{OP01} which allows to avoid ultraviolet divergences). The
Hamiltonian (\ref{H}) leads to the following equations of motion for the
annihilation operators:
\begin{eqnarray}
&&i \dot{\hat{\psi }}\left( {\bf p},t\right) = \epsilon \left( {\bf
 p}\right) \hat{\psi }\left( {\bf p},t\right)  \nonumber
\\
&&+{1\over {\cal V}}\sum\limits^{}_{{\bf q},{\bf p}^\prime }U\left(
 q\right) \hat{\psi }^{\dag }\left( {\bf p}^\prime ,t\right)
 \hat{\psi }\left( {\bf p}-{\bf q},t\right) \hat{\psi }\left( {\bf
 p}^\prime +{\bf q},t\right)  \label{pside}
\end{eqnarray}
Following a general method \cite{PB96,PBS98,G96,WWCH99} let us separate the
condensate part by a projection onto a coherent state $|\Phi \rangle
 $ defined so
that
\begin{equation}
\hat{\psi }\left( {\bf p},t\right) |\Phi \rangle =\delta _{{\bf
 p}0}\bar{\Phi }\left( t\right) |\Phi \rangle  ,\qquad \langle \Phi
|\Phi \rangle =1 ,
\end{equation}
where the mean field $\bar{\Phi }\left( t\right) =\langle \hat{\psi
 }\left( {\bf p},t\right) \rangle =\langle \Phi |\hat{\psi }\left(
 {\bf p},t\right) |\Phi \rangle $ describes the
ground-state condensate occupying the momentum state ${\bf p}=0$. The
annihilation operator can be written as
\begin{equation}
\hat{\psi }\left( {\bf p},t\right) =\left\lbrack \Phi \left( t\right)
 |\Phi \rangle \langle \Phi |+\hat{\xi }\left( {\bf p},t\right)
 \right\rbrack \exp\left( -i U\left( 0\right) N t/{\cal V}\right)  ,
 \label{psiphi}
\end{equation}
where $\Phi \left( t\right) =\bar{\Phi }\left( t\right) \exp\left( i
 U\left( 0\right) N t/{\cal V}\right) $, the operators $\hat{\xi
 }\left( {\bf p},t\right) $ represent
the non-condensate part (fluctuations) having the zero expectation
value,
\begin{equation}
\langle \hat{\xi }\left( {\bf p},t\right) \rangle =0\text{,
 }\label{csiev}
\end{equation}
and
\begin{eqnarray}
N&&=\sum\limits^{}_{{\bf p}}\langle \hat{\psi }^{\dag }\left( {\bf
 p},t\right) \hat{\psi }\left( {\bf p},t\right) \rangle  \nonumber
\\
&&=|\Phi \left( t\right) |^{2}+\sum\limits^{}_{{\bf p}}\langle
 \hat{\xi }^{\dag }\left( {\bf p},t\right) \hat{\xi }\left( {\bf
 p},t\right) \rangle \label{Nbos}
\end{eqnarray}
is the conserved total number of bosons. The exponential
factor in Eq.\ (\ref{psiphi}) shifts the zero energy to the energy
of non-excited condensate.

Substitution of Eq.\ (\ref{psiphi}) into Eq.\ (\ref{pside}),
combined with the property (\ref{csiev}), produces equations of
motion for $\Phi \left( t\right) $ and $\hat{\xi }\left( {\bf
 p},t\right) $. Assuming a small occupation of the
fluctuations, let us keep in these equations only terms of lowest
non-vanishing order in $\hat{\xi }\left( {\bf p},t\right) $ and
 $\hat{\xi }^{\dag }\left( {\bf p},t\right) $, obtaining the following
equations:
\begin{eqnarray}
i\dot{\Phi }\left( t\right) =&&{1\over {\cal V}}\sum\limits^{}_{{\bf
 p}}U\left( p\right) \langle \hat{\xi }\left( {\bf p},t\right)
 \hat{\xi }\left( -{\bf p},t\right) \rangle \Phi ^{*}\left( t\right)
 \nonumber
\\
&&+{1\over {\cal V}}\sum\limits^{}_{{\bf p}}U\left( p\right) \langle
 \hat{\xi }^{\dag }\left( {\bf p},t\right) \hat{\xi }\left( {\bf
 p},t\right) \rangle \Phi \left( t\right)  \label{phi}
\\
i\dot{\hat{\xi }}\left( {\bf p},t\right) &&=d\left( p\right) \hat{\xi
 }\left( {\bf p},t\right) +g\left( p\right) \hat{\xi }^{\dag }\left(
-{\bf p},t\right)  , \label{csip}
\end{eqnarray}
where
\begin{equation}
d\left( p\right) =\epsilon \left( p\right) +U\left( p\right) n_{0}
 ,\qquad g\left( p\right) =U\left( p\right) \Phi ^{2}/{\cal V} ,
\end{equation}
and $n_{0}=|\Phi |^{2}/{\cal V}$  is the condensate density. This
 approximate
method, that leads to linear operator equations for the fluctuations, is
similar to the parametric approximation used in quantum optics (see
Ref.\ \cite{SZ97}). It has been applied to the study of a molecular BEC
dissociation
in Refs.\ \cite{VYA01,YBJ01}), and is valid as long as the depletion of the
condensate is negligibly small. An applicability criterion for
this method  is
presented in Sec.\ \ref{CondDepl} below [see Eq.\ (\ref{ValRan})].

The present approximation is gapless [see Eq.\ (\ref{lambda}) below]
and, therefore, non-conserving (see Ref.\ \cite{G96}), although the
average number of the atoms is conserved [see Eq.\ (\ref{Nbos}) above].

\subsection{Stable and unstable modes} \label{StabUnstMod}

Equation (\ref{csip}) combined with the corresponding
equation for $\hat{\xi }^{\dag }\left( -{\bf p},t\right) $ provide a
 set of two linear operator equations
(similar to the set considered in Refs.\ \cite{VYA01,YBJ01}). This set
can be solved exactly, using common methods reducing it to an
eigenproblem, by assuming the time-independence of $d\left( p\right) $
 and $g\left( p\right) $. This
assumption is corroborated by Eq.\ (\ref{phi}), which shows that  the
variation of $\Phi $  is of the second order in $\hat{\xi }\left( {\bf
 p},t\right) $ and therefore can be
neglected. Indeed, the validity of the parametric approximation
requires a small variation of the condensate density, and the
dominant part of the condensate phase is accumulated in the
exponential factor in Eq.\ (\ref{psiphi}).

As  in the case considered in Ref.\ \cite{VYA01}, the form
of the solution depends on the ratio of the detuning $d\left(
 p\right) $ to the
coupling strength $g\left( p\right) $. All momentum modes observing
\begin{equation}
|g\left( p\right) |>d\left( p\right) \text{,  or     }\epsilon \left(
 p\right) +2U\left( p\right) n_{0}<0 \label{inst}
\end{equation}
are unstable,  and the corresponding solutions attain the aperiodic form
\begin{eqnarray}
\hat{\xi }\left( {\bf p},t\right) =&&\left\lbrack \cosh \left(
 \lambda \left( p\right) t\right)  -i {d\left( p\right) \over \lambda
 \left( p\right) }\sinh \left( \lambda \left( p\right) t\right)
 \right\rbrack \hat{\xi }\left( {\bf p},0\right)  \nonumber
\\
&&-i{g\left( p\right) \over \lambda \left( p\right) }\sinh \left(
 \lambda \left( p\right) t\right)  \hat{\xi }^{\dag }\left( -{\bf
 p},0\right)  , \label{solap}
\end{eqnarray}
where
\begin{eqnarray}
\lambda \left( p\right) &&=\sqrt{|g\left( p\right) |^{2}-d^{2}\left(
 p\right) } \nonumber
\\
&&=\sqrt{\epsilon \left( p\right) \left( -2U\left( p\right) n_{0}
-\epsilon \left( p\right) \right) } . \label{lambda}
\end{eqnarray}
The condition (\ref{inst}) can be obeyed only when $U\left( p\right)
<0$, i. e.,
with a negative scattering length, and for excitations with an energy
$\epsilon \left( p\right) <2|U\left( p\right) |n_{0}$. Otherwise, whenever
 $|g\left( p\right) |<d\left( p\right) $, the modes are stable and the
solutions are oscillatory functions of the form
\begin{eqnarray}
\hat{\xi }\left( {\bf p},t\right) =&&\left\lbrack \cos \left(
|\lambda \left( p\right) |t\right)  -i {d\left( p\right) \over
|\lambda \left( p\right) |}\sin \left( |\lambda \left( p\right)
|t\right) \right\rbrack \hat{\xi }\left( {\bf p},0\right)  \nonumber
\\
&&-i{g\left( p\right) \over |\lambda \left( p\right) |}\sin \left(
|\lambda \left( p\right) |t\right)  \hat{\xi }^{\dag }\left(- {\bf
 p},0\right)  . \label{solp}
\end{eqnarray}

In a case of intervening mode ($\lambda(p)=0$) the solution is
a linear function of time
\begin{equation}
\hat{\xi }\left( {\bf p},t\right) =\left\lbrack 1-i d(p) t
\right\rbrack \hat{\xi }\left( {\bf p},0\right)
-i g(p) t  \hat{\xi }^{\dag }\left(- {\bf p},0\right)  .
\label{soli}
\end{equation}
This case can take place for the spurious (Goldstowne) mode with
a vanishing excitation energy ($\epsilon(p)=0$, see Refs.\
\cite{BR86,LY96}), or at $\epsilon(p)=2|U(p)|n_0$.

\subsection{Regular and inverted oscillators} \label{RegInvOsc}

The stable modes can be treated by the Bogolubov transformation.
It is therefore instructive to establish a connection between the present
and the Bogolubov methods. The equations of motion (\ref{csip}) can be
derived from the following Hamiltonian
constituent involving two counter-propagating modes,
\begin{eqnarray}
&&\hat{H}\left( {\bf p}\right) =d\left( p\right) \left\lbrack
 \hat{\xi }^{\dag }\left( {\bf p},t\right) \hat{\xi }\left( {\bf
 p},t\right) +\hat{\xi }^{\dag }\left( -{\bf p},t\right) \hat{\xi
 }\left( -{\bf p},t\right) \right\rbrack  \nonumber
\\
&&+g^{*}\left( p\right) \hat{\xi }\left( {\bf p},t\right) \hat{\xi
 }\left( -{\bf p},t\right) +g\left( p\right) \hat{\xi }^{\dag }\left(
 {\bf p},t\right) \hat{\xi }^{\dag }\left( -{\bf p},t\right)  .
\label{Hp}
\end{eqnarray}
The transformation
\begin{equation}
\hat{\xi }_{1,2}\left( {\bf p},t\right) =2^{-1/2}\left( \hat{\xi
 }\left( {\bf p},t\right) \pm \hat{\xi }\left( -{\bf p},t\right)
 \right) \exp\left( -{i\over 2}\arg g\left( p\right) \right)
 \label{csi12}
\end{equation}
introduces a new pair of field operators,
 $\hat{\xi }_{1}\left( {\bf p},t\right) $  and
 $\hat{\xi }_{2}\left( {\bf p},t\right) $, whose equations of motion
are decoupled by separating the Hamiltonian into
$\hat{H}\left( {\bf p}\right) =\hat{H}_{1}\left( {\bf p}\right)
+\hat{H}_{2}\left( {\bf p}\right) $, where
\begin{equation}
\hat{H}_{j}=d\hat{\xi }^{\dag }_{j}\hat{\xi }_{j}-{1\over 2}\left(
-1\right) ^{j}|g|\left( \hat{\xi }^{\dag }_{j}\hat{\xi }^{\dag }_{j}
+\hat{\xi }_{j}\hat{\xi }_{j}\right), \quad  j=1,2  . \label{Hj}
\end{equation}
(Hereafter in this subsection the arguments ${\bf p}$ and $t$ are
 dropped out, for simplicity's sake.)
A system described by the Hamiltonian (\ref{Hj}) was
solved in Ref.\ \cite{VYA01}, and the solutions have a form similar to
Eq.\ (\ref{solap}), or Eq.\ (\ref{solp}), according to the
ratio of $d$ to $g$.

Let us introduce canonically conjugate operators for the
coordinates $\hat{Q}_{j}$  and the momenta $\hat{P}_{j}$, such that
 $\hat{\xi }_{j}=2^{-1/2}\left( \hat{Q}_{j}+i \hat{P}_{j}\right) $,
writing out the Hamiltonian (\ref{Hj}) as
\begin{equation}
\hat{H}_{j}={d+\left( -1\right) ^{j}|g|\over 2}\hat{P}^{2}_{j}+{d
-\left( -1\right) ^{j}|g|\over 2}\hat{Q}^{2}_{j}-{d\over 2} .
 \label{HPQ}
\end{equation}
If $|g|<d$, this Hamiltonian describes a harmonic oscillator and can be
transformed into the standard form representing free Bogolubov
quasiparticles [the total transformation, incorporating (\ref{csi12}),
being indeed the Bogolubov transformation]. The motion of the harmonic
oscillator is periodic, in agreement with the periodic behavior of the
solution  (\ref{solp}).

A different situation takes place if the condition (\ref{inst})
holds. In this case one of the coefficients (preceding $\hat{P}^{2}_{j}$
  or $\hat{Q}^{2}_{j}$) is
negative, and the  Hamiltonian (\ref{HPQ}) represents an inverted
oscillator. It cannot be transformed to a form representing
free quasiparticles,
and therefore a Bogolubov transformation cannot exist, in
agreement with Ref.\ \cite{BB94}. The infinite aperiodic
motion of the inverted oscillator is exhibited in Eq.\ (\ref{solap})
and in the instability of the mode.

In a case of intervening mode ($\lambda(p)=0$) Eq.\ (\ref{lambda})
gives $d(p)=|g(p)|$, one of the coefficients (preceding
$\hat{Q}^{2}_{j}$  or $\hat{P}^{2}_{j}$) vanishes, and the Hamiltonian
(\ref{HPQ}) represents a motion without restoring forces or an oscillator
with infinitely large mass, respectively (see Ref.\ \cite{BR86}). Such
systems demonstrate linear time-dependence of $\hat{P}_{j}$  or
$\hat{Q}_{j}$, respectively, in agreement with the solution (\ref{soli}).
The Bogolubov transformation does not exist in this case as well. In
the case of the Goldstowne mode ($p=0$) the Hamiltonian (\ref{Hp})
has already the form (\ref{Hj}), and the transformation (\ref{csi12})
is not necessary. This mode brings only one term $\hat{Q}^{2}(0,t)$
into the total Hamiltonian, in agreement with Ref.\ \cite{LY96}
(up to the exchange of the coordinate and the momentum).

\subsection{Dynamics of instabilities} \label{InstDyn}

Consider the dynamics of fluctuations in an unstable BEC, starting
from the state of a pure condensate
\begin{equation}
\langle \hat{\xi }^{\dag }\left( {\bf p},0\right) \hat{\xi }\left(
 {\bf p},0\right) \rangle =\langle \hat{\xi }\left( {\bf p},0\right)
 \hat{\xi }\left( -{\bf p},0\right) \rangle =0 . \label{PureCond}
\end{equation}
This problem is not only of a speculative interest;
it actually describes
the recent JILA experiment \cite{D01}, in which the elastic scattering
length was varied very fast from the zero value (of the ideal condensate) to
a negative value (of an unstable condensate). Given the initial conditions
(\ref{PureCond}), Eq.\ (\ref{solap}) leads to the following dynamics for the
occupation of the unstable modes,
\begin{equation}
\langle \hat{\xi }^{\dag }\left( {\bf p},t\right) \hat{\xi }\left(
 {\bf p},t\right) \rangle ={|g\left( p\right) |{ } ^{2}\over \lambda
 ^{2}\left( p\right) }\sinh^{2}\left( \lambda \left( p\right) t\right
)  ,\label{den}
\end{equation}
and for the two-particle correlations,
\begin{eqnarray}
\langle \hat{\xi }\left( {\bf p},t\right) \hat{\xi }\left( -{\bf
 p},t\right) \rangle =&&-i{g\left( p\right) \over \lambda \left(
 p\right) } \cosh \left( \lambda \left( p\right) t\right)  \sinh
 \left( \lambda \left( p\right) t\right)  \nonumber
\\
&&-{d\left( p\right) g\left( p\right) \over \lambda ^{2}\left(
 p\right) }\sinh^{2}\left( \lambda \left( p\right) t\right)  .
 \label{cor}
\end{eqnarray}
It is not surprising that this dynamics is similar to the
dynamics of spontaneous dissociation of a molecular condensate,
studied in Ref.\ \cite{VYA01}, as the non-condensate modes can be
described in terms of the fields $\hat{\xi }_{1,2}\left( {\bf
 p},t\right) $ [see Eq.\ (\ref{csi12})] by
the Hamiltonian (\ref{Hj}), which is equivalent to the one considered in
Ref.\ \cite{VYA01}. By analogy to the results of Ref.\ \cite{VYA01},
the non-condensate atoms are formed in squeezed states
involving atom pairs of opposite momenta. These states are similar to
the two-mode squeezed states of fluctuations in a stable BEC studied in
Ref.\ \cite{RGB01}. The states are perfectly squeezed and have zero
variance in the relative number of particles with momenta ${\bf p}$
 and $-{\bf p}$,
being optimally entangled in the entropic sense. The uncertainty of
the generalized coordinates
\begin{equation}
\hat{X}_{j}\left( {\bf p},t\right) =2^{-1/2}\left\lbrack \hat{\xi
 }_{j}\left( {\bf p},t\right) \exp\left( i\theta _{j}\right) +
 \hat{\xi }^{\dag }_{j}\left( {\bf p},t\right) \exp\left( -i\theta
 _{j}\right) \right\rbrack  ,
\end{equation}
where the operators $\hat{\xi }_{j}\left( {\bf p},t\right) $
(with $j=1,2$) are defined by Eq.\ (\ref{csi12}), attains
the maximal and minimal values
\begin{eqnarray}
\langle \hat{X}^{2}_{j}\left( {\bf p},t\right) \rangle _{\pm
 }&&={1\over 2}\Biggl|\left\lbrack \cosh^{2}\left( \lambda \left(
 p\right) t\right) +{d^{2}\left( p\right) \over \lambda ^{2}\left(
 p\right) }\sinh^{2}\left( \lambda \left( p\right) t\right)
 \right\rbrack ^{1/2} \nonumber
\\
&&\pm {g\left( p\right) \over \lambda \left( p\right) } \sinh \left(
 \lambda \left( p\right) t\right) \Biggr|^{2}\mathrel{ \mathop \approx
 _{\lambda t \gg 1}} \cases{{|g\left( p\right) |{ } ^{2}\over
 2\lambda ^{2}\left( p\right) }e^{2\lambda t}\cr {\lambda ^{2}\left(
 p\right) \over 2|g\left( p\right) |{ } ^{2}}e^{-2\lambda t}}
\end{eqnarray}
at the two orthogonal values of the respective phase angles $\theta _{j}$,
\begin{equation}
\theta _{j\pm }=\pm \left( -1\right) ^{j}{\pi \over 4}+{1\over
 2}\arctan \left\lbrack {d\left( p\right) \over \lambda \left(
 p\right) }\tanh \left( \lambda \left( p\right) t\right)
 \right\rbrack  .
\end{equation}

A common method for studying BEC stability  is the Liapunov
analysis (see e.\ g.\ Ref.\ \cite{DGPS99}). This method produces
equations for small perturbations $\zeta \left( {\bf p},t\right) $ to
 the mean field. These
equations are equivalent to Eq.\ (\ref{csip}), in which the operators
$\hat{\xi }\left( {\bf p},t\right) $ are replaced by $c$-number
 functions $\zeta \left( {\bf p},t\right) $. In the case
studied here one obtains for the Liapunov exponents the expression
$\lambda $ given by Eq.\
(\ref{lambda}). This type of analysis attributes an exponential
gain to the small
perturbation, $|\zeta \left( {\bf p},t\right) |^{2}\sim \exp\left(
 2\lambda t\right) $, when the condition  (\ref{inst}) is
obeyed. This result is in qualitative agreement  with the quantum
analysis used here, which attributes an even faster
(super-exponential) gain to the
fluctuations [see Eq.\ (\ref{den})], as
\begin{equation}
{d\over dt}\ln\left( \langle \hat{\xi }^{\dag }\left( {\bf p},t\right
) \hat{\xi }\left( {\bf p},t\right) \rangle \right) =2\lambda \left(
 p\right) \coth\lambda \left( p\right) t>2\lambda \left( p\right)  .
\end{equation}
The dynamics of the fluctuations has been studied in Ref.\
\cite{VYA01}. It is shown there that the quantum (squeezed) state
always grows faster than a classical (coherent) state. Therefore the
Liapunov analysis concerning classical fields underestimates
the fluctuation growth, becoming correct only when $\lambda t\gg 1$,
 where the
fluctuations are large enough to be considered as classical ones.

A formal application of the Bogolubov transformation at $U<0$ (although
incorrect, as had been observed by Bogolubov himself; see Ref.\
\cite{BB94}) produces quasiparticles with imaginary energies
 $i\lambda $, which
correspond to  exponential growth. However, the transformed field
operators do not obey the proper Bose commutation rules.

The homogeneous boson gas with a negative scattering length considered
above always has unstable modes due to the continuous nature of the
excitation spectrum.
The situation changes in the case of a trapped gas having a discrete
excitation spectrum. Following Ref.\ \cite{D96}, consider $\lambda
 \left( p\right) $ as a
function of the condensate density $n_{0}$  for a fixed value of
 $p=p_{m}$
chosen so that $\epsilon \left( p_{m}\right) =\epsilon _{m}$, where
 $\epsilon _{m}$  is the energy of an excited state
of the trapped condensate. At low condensate density $\lambda \left(
 p_{m}\right) $ is imaginary,
corresponding to the well-known stability of a trapped condensate with
a negative scattering length at low occupations (see Ref.\
\cite{DGPS99}). At $n_{0}>{1\over 2}\epsilon _{m}/|U\left(
 p_{0}\right) |$ the value of $\lambda \left( p_{m}\right) $ becomes
 real,
and therefore describes an instability of the condensate
(leading to its collapse). Although the
present approach does not take into account the inhomogeneity of the
condensate and the variation of its size, the dependence of $\lambda
 \left( p_{m}\right) $ on
$n_{0}$  [see Eq.\ (\ref{lambda})] is in  qualitative agreement with
 the one
obtained from the numerical calculations of Ref.\ \cite{D96} for the
monopole mode.

\subsection{Dynamics of the condensate} \label{CondDepl}

Let us consider now the dynamics of  condensate depletion. In
the limit ${\cal V}\rightarrow \infty $, Eqs.\ (\ref{den}), (\ref{cor}
), and  (\ref{phi}) lead to
the following equation for the  condensate density,
\begin{equation}
\dot{n}_{0}=-{m p_{s}U^{2}n^2_0\over 2\pi { } ^{2}}\left(
 I_{\text{unst}}+I_{\text{stab}}\right)  ,
\end{equation}
where
\begin{eqnarray}
&&I_{\text{unst}}={1\over m p_s}\int\limits^{p{ }
 _{s}}_{0}p^{2}{\sinh\left( 2\lambda \left( p\right) t\right) \over
 \lambda \left( p\right) }d p \nonumber
\\
&&
\\
&&I_{\text{stab}}={1\over m p_s}\int\limits^{\infty }_{p{ }
 _{s}}p^{2}{\sin\left( 2|\lambda \left( p\right) |t\right) \over
|\lambda \left( p\right) |}d p \nonumber
\end{eqnarray}
are the contributions of the unstable ($p<p_{s}$) and stable ($p
>p_{s}$)
modes, respectively,
\begin{equation}
p_{s}=\sqrt{4m|U|n{ } _{0}}
\end{equation}
is the boundary between the stable and unstable modes, and the
contact potential (\ref{Ua}) is used hereafter in this section.
The statistical weight of intervening mode with $\lambda(p)=0$
is negligible  compared to the continuums of the stable and unstable
modes. Unlike a case of stable condensate, where the unrestricted
behavior of the Goldstowne mode requires a special treatment
(see Ref.\ \cite{LY96}), in the present case the effect of
linearly-growing intervening modes is negligible compared to the
effect of exponentially-growing unstable modes.

\begin{figure}
\psfig{figure=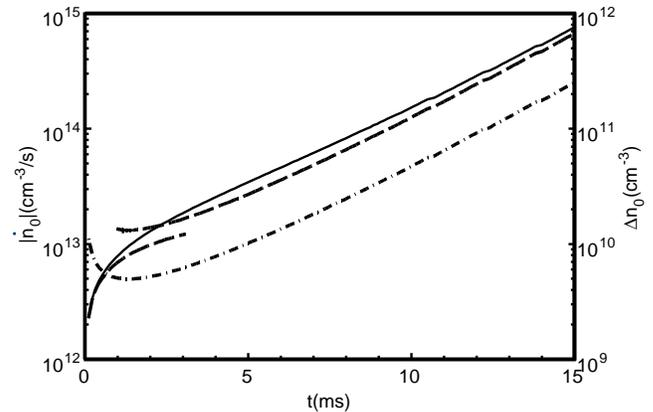,width=3.375in}

\caption{Time-dependence of the loss of condensate density $\Delta n_0$
under conditions of the ${}^{85}$Rb JILA  experiment \protect\cite{D01}
($n_{0}\approx 10^{12}$cm$^{-3}$). The results of numerical integration
(solid line) are compared to the approximate equations
(\protect\ref{Deltann}) and (\protect\ref{Deltane}) (dashed line, left
and right segments, respectivelly). The dot-dashed line represents
the condensate depletion rate $\dot{n}_{0}$ given by
Eq.\ (\protect\ref{ndot}). \label{Fg}}

\end{figure}

Substitution of $p=p_{s}\cos{\theta \over 2}$ allows us to express
 $I_{\text{unst}}$  in terms of the
Weber functions $E_{\nu }\left( z\right) $ (see Ref.\ \cite{BE53}):
\begin{eqnarray}
I_{\text{unst}}&&=\int\limits^{\pi }_{0}\cos{\theta \over 2}
 \sinh\left( \tau  \sin\theta \right) d\theta  \nonumber
\\
&&=i{\pi \over 2}\left\lbrack E_{1/2}\left( i\tau \right) +E_{-1
/2}\left( i\tau \right) \right\rbrack  , \label{Iinst}
\end{eqnarray}
where $\tau =2t/t_{\text{NL}}$,  and $t_{\text{NL}}=\left( |U|n_{0}\right) ^{-1}$
 is the nonlinear interaction
time (see Ref.\ \cite{TBJ00}). Approximate expressions for
 $I_{\text{unst}}$,
holding at small and large values of $t$, have the following respective forms,
\begin{eqnarray}
&&I_{\text{unst}}\mathrel{ \mathop \approx_{\tau \rightarrow 0}}{\pi
 \over 2}{\tau \over \sqrt{2}\Gamma \left( 7/4\right) \Gamma \left( 5
/4\right) } \nonumber
\\
&& \label{Iinsta}
\\
&&I_{\text{unst}}\mathrel{ \mathop \approx_{ \tau \rightarrow
 \infty}}{1\over 2}\sqrt{{\pi \over \tau }}e^{\tau }-{3\over 16\tau } .
 \nonumber
\end{eqnarray}
The contribution of the stable modes can be evaluated by
substituting $p=p_{s}\cosh{\theta \over 2}$, as
a result of which
\begin{eqnarray}
I_{\text{stab}}&&=\int\limits^{\infty }_{0}\cosh{\theta \over 2}
 \sin\left( \tau  \sinh\theta \right) d\theta  \nonumber
\\
&&={1\over 2}\sqrt{{\pi \over \tau }}e^{\tau }-i{\pi \over
 2}\left\lbrack E_{1/2}\left( i\tau \right) +E_{-1/2}\left( i\tau
 \right) \right\rbrack . \label{Istab}
\end{eqnarray}
Its limiting values can be approximately expressed as
\begin{eqnarray}
&&I_{\text{stab}}\mathrel{ \mathop \approx_{ \tau \rightarrow0}}{\pi
 \over 2}\left(  {1\over \sqrt{\pi \tau }}+\sqrt{{\tau \over \pi }} -
 {\tau \over \sqrt{2}\Gamma \left( 7/4\right) \Gamma \left( 5/4\right
) }\right)  \nonumber
\\
&& \label{Istaba}
\\
&&I_{\text{stab}}\mathrel{ \mathop \approx_{ \tau \rightarrow
 \infty}}{3\over 16\tau } . \nonumber
\end{eqnarray}
Finally, the rate of  condensate depletion can be expressed as
\begin{eqnarray}
\dot{n}_{0}&&=-{m p_{s}U^{2}n^2_0\over 4\pi \sqrt{\pi \tau }}e^{\tau
 } \nonumber
\\
&&=-4\sqrt{{2\pi \over mt}}a^{2}_{a}n^{2}_{0}\exp\left( 8\pi |a_{a}
|n_{0}t/m\right) , \label{ndot}
\end{eqnarray}
and the loss of  condensate density attains the limiting forms
\begin{mathletters} \label{Deltan} \begin{eqnarray}
\Delta n_{0}&&\approx  4 n_{0}\sqrt{|a_{a}|^{3}n_{0}\tau }
 =8\sqrt{{2\pi t\over m}}a^{2}_{a}n^{2}_{0},\qquad t\ll t_{\text{NL}}
\label{Deltann}
\\
\Delta n_{0}&&\approx 2 e^{\tau }n_{0}\sqrt{|a_{a}|^{3}n_{0}/\tau }
 \nonumber
\\
&&=\sqrt{{m\over 2\pi t}}a_{a}n_{0}\exp\left( 8\pi |a_{a}|n_{0}t
/m\right) ,\qquad t\gg t_{\text{NL}} \label{Deltane}
\end{eqnarray} \end{mathletters}
at the two distinguishable loss regimes.
At the regime of non-stationarity $t\ll t_{\text{NL}}$  the depletion is
mostly due to an occupation of
 the stable modes, which carry  much more  statistical weight than the
unstable modes [$I_{\text{stab}}\gg I_{\text{unst}}$, see Eqs.\
(\ref{Iinsta}) and
(\ref{Istaba})]. The statistical weight is infinitely large,
 leading to
a singularity in the depletion rate (\ref{ndot}), at $t=0$. This
singularity is absent in Eq.\ (\ref{Deltan}), and  can be eliminated
by using of an interatomic interaction with a finite range in place of
the contact potential (\ref{Ua}). At $t>0$ the occupation of the modes
with $p>\sqrt{m/t}$ begins to decrease due to their oscillating
 behavior,
leading to decrease of the depletion rate. At the regime of Bose-Enhancement
$t\gg t_{\text{NL}}$  the
depletion is dominated by the exponentially increasing occupation of the
unstable modes [$I_{\text{stab}}\ll I_{\text{unst}}$, see Eqs.\
(\ref{Iinsta}) and
(\ref{Istaba})]. The approximate expressions (\ref{Deltan}) are
compared in Fig.\ \ref{Fg} to results of numerical integration of
Eq.\ (\ref{ndot}).

One should be reminded that the parametric approximation, on which all
the analysis above
rests, is applicable only as long as $\Delta n_{0}\ll n_{0}$.
In the case of one
 excited mode
\cite{VYA01}, this approximation is in good agreement with numerical
results while $\Delta n_{0}/n_{0}<0.2$. In the case of a dilute
 condensate ($|a_{a}|^{3}n_{0}\ll 1$,
see Ref.\ \cite{DGPS99}), the validity range
\begin{equation}
t/t_{\text{NL}}<{1\over 4}|\ln\left( |a_{a}|^{3}n_{0}\right) | \label{ValRan}
\end{equation}
includes a region $t>t_{\text{NL}}$  where the exponential Bose-enhancement
factor can reach a rather large value, of the order of $\left( |a_{a}
|^{3}n_{0}\right) ^{-1/2}$.

Equation (\ref{Deltann}) predicts small depletion at times $t<t_{\text{NL}}$
(with $\Delta n_{0}/n_{0}<4\sqrt{|a_{a}|^{3}n{ } _{0}}\ll 1$, given the
 dilution parameter $|a_{a}|^{3}n_{0}$  is small).
Therefore, the development of instabilities in BEC, including collapse,
starts with a delay time $t\sim t_{\text{NL}}$  after turning on the
 interatomic
interactions, although the collapse itself is a nonlinear phenomenon
that cannot be treated by the parametric approximation. Under the conditions
of the ${}^{85}$Rb  experiment \cite{D01}
(where $n_{0}\approx 10^{12}$cm$^{-3}$ and $a_{a}\approx -20$nm), the
expected delay is
$t_{\text{NL}}\approx 5$ms (cf.\ Fig.\ \ref{Fg}), in  agreement with the
experimentally observed delay time
for the collapse. In this case the parametric approximation is valid
as long as $t<3t_{\text{NL}}$, allowing for values of the exponential Bose-enhancement
factor in Eq.\ (\ref{Deltane}) exceeding 100. It should be noted that
the condensate dynamics can be sensitive to initial
correlations and the spatial inhomogeneity neglected here.

\section{Collisions of two condensates} \label{twocond}

\subsection{Separation of condensates and fluctuations}
\label{TCSep}

The approach developed in Sec.\ \ref{Ground} above can also be
applied to a collision of two free counterpropagating BEC waves with momenta
$\pm {\bf p}_{0}$  (see Ref.\ \cite{BTBJ00}). In this case the
 annihilation operator
should be represented as
\begin{eqnarray}
\hat{\psi }\left( {\bf p},t\right) =\Biggl\lbrack &&\Phi _{+}\left(
 t\right) |\Phi _{+}\rangle \langle \Phi _{+}|\exp\left( -2i S_{
-}\left( t\right) \right)  \nonumber
\\
&&+\Phi _{-}\left( t\right) |\Phi _{-}\rangle \langle \Phi _{-}
|\exp\left( -2i S_{+}\left( t\right) \right)  \nonumber
\\
&&+\hat{\xi }\left( {\bf p},t\right) \exp\left( -i S_{+}\left(
 t\right) -i S_{-}\left( t\right) \right) \Biggr\rbrack  \label{psi2p}
\end{eqnarray}
where
\begin{equation}
S_{\pm }={U\left( 2p_{0}\right) \over 2{\cal V}} \int\limits^{t}_{0}
|\Phi _{\pm }\left( t^\prime \right) |^{2}dt^\prime +{U\left( 0\right
) \over 2{\cal V}}N t+{1\over 2}\epsilon \left( p_{0}\right) t .
\end{equation}
Here $\Phi _{\pm }\left( t\right) $ describe the two condensate fields with
 momenta $\pm {\bf p}_{0}$,
respectively, the coherent states $|\Phi _{\pm }\rangle $ are defined
 by
\begin{eqnarray}
&&\hat{\psi }\left( {\bf p},t\right) |\Phi _{\pm }\rangle =\Phi _{\pm
 }\left( t\right) \exp\left( -2i S_{\mp }\left( t\right) \right)
 \delta _{\pm {\bf p}_0{\bf p}}|\Phi _{\pm }\rangle  \nonumber
\\
&&
\\
&&\langle \Phi _{\pm }|\Phi _{\pm }\rangle =1 , \nonumber
\end{eqnarray}
and the conserved total number of atoms is
\begin{equation}
N=|\Phi _{+}\left( t\right) |^{2}+|\Phi _{-}\left( t\right) |^{2}
+\sum\limits^{}_{{\bf p}}\langle \hat{\xi }^{\dag }\left( {\bf
 p},t\right) \hat{\xi }\left( {\bf p},t\right) \rangle .
\end{equation}
 Substitution of Eq.\ (\ref{psi2p}) into Eq.\
(\ref{pside}) gives the following equations of motion for the
fluctuations,
\begin{equation}
i\dot{\hat{\xi }}\left( {\bf p},t\right) =d_{2}\left( p\right)
 \hat{\xi }\left( {\bf p},t\right) +g_{2}\left( p\right) \hat{\xi
 }^{\dag }\left( -{\bf p},t\right) +D , \label{csip2}
\end{equation}
where
\begin{eqnarray}
d_{2}\left( p\right) =&&\epsilon \left( p\right) -\epsilon \left(
 {\bf p}_{0}\right) +\left\lbrack U\left( |{\bf p}-{\bf p}_{0}|\right
) -{1\over 2}U\left( 2p_{0}\right) \right\rbrack n_{+} \nonumber
\\
&&+\left\lbrack U\left( |{\bf p}+{\bf p}_{0}|\right) -{1\over
 2}U\left( 2p_{0}\right) \right\rbrack n_{-} \label{d2}
\\
g_{2}\left( p\right) &&=\left\lbrack U\left( |{\bf p}-{\bf p}_{0}
|\right) +U\left( |{\bf p}+{\bf p}_{0}|\right) \right\rbrack \Phi _{
+}\Phi _{-}/{\cal V} , \label{g2}
\end{eqnarray}
and $n_{\pm }=|\Phi _{\pm }|^{2}/{\cal V}$ are the densities of the
 condensate waves. The
term
\begin{eqnarray}
D=\{&&\Phi ^{*}_{-}\left( t\right) \Phi _{+}\left( t\right) \left[
 U\left( 2{\bf p}_{0}\right) +U\left( |{\bf p}-{\bf p}_{0}|\right)
 \right] \hat{\xi }\left( {\bf p}-2{\bf p}_{0},t\right)  \nonumber
\\
&&+\Phi ^{2}_{+}\left( t\right) U\left( |{\bf p}-{\bf p}_{0}|\right)
 \hat{\xi }^{\dag }\left( -{\bf p}+2{\bf p}_{0},t\right) \} \nonumber
\\
\times&&\exp\{2i[S_+(t)-S_-(t)]\}  \nonumber
\\
+\{&&\Phi ^{*}_{+}\left( t\right) \Phi _{-}\left( t\right) \left[
 U\left( 2{\bf p}_{0}\right) +U\left( |{\bf p}+{\bf p}_{0}|\right)
 \right] \hat{\xi }\left( {\bf p}+2{\bf p}_{0},t\right)  \nonumber
\\
&&+\Phi ^{2}_{-}\left( t\right) U\left( |{\bf p}+{\bf p}_{0}|\right)
 \hat{\xi }^{\dag }\left( -{\bf p}-2{\bf p}_{0},t\right) \} \nonumber
\\
\times&&\exp\{2i[S_-(t)-S_+(t)]\}
\end{eqnarray}
describes couplings of fluctuations with momenta different by
$2{\bf p}_{0}$. Whenever
\begin{equation}
p^{2}_{0}\gg 4 m U\sqrt{n_{+}n{ } _{-}} \label{highp}
\end{equation}
the disparity of the energies of such fluctuations far exceeds
the coupling  strengths, and the term $D$ can be neglected
in Eq.\ (\ref{csip2}).
This allows us also to neglect correlations of such fluctuations in
the equations of motion for the condensate fields, which then attain the
 form
\begin{eqnarray}
i\dot{\Phi }_{\pm }\left( t\right) =&&{1\over {\cal
 V}}\sum\limits^{}_{{\bf p}}U\left( |{\bf p}-{\bf p}_{0}|\right)
 \langle \hat{\xi }\left( {\bf p},t\right) \hat{\xi }\left( -{\bf
 p},t\right) \rangle \Phi ^{*}_{\mp }\left( t\right)  \nonumber
\\
&&+{1\over {\cal V}}\sum\limits^{}_{{\bf p}}U\left( |{\bf p}-{\bf
 p}_{0}|\right) \langle \hat{\xi }^{\dag }\left( {\bf p},t\right)
 \hat{\xi }\left( {\bf p},t\right) \rangle \Phi _{\pm }\left( t\right
)  \label{phi2}
\end{eqnarray}
Equation (\ref{csip2}) with $D$ neglected has solutions of the form Eqs.\
(\ref{solap}), (\ref{solp}), or (\ref{soli}), according to the ratio of
$|g_{2}|$ to $d_{2}$, in which $g$, $d$, and $\lambda $ are
respectively replaced by $g_{2}$,
 $d_{2}$, and
\begin{equation}
\lambda _{2}\left( p\right) =\sqrt{|g_{2}\left( p\right) |^{2}
-d^{2}_{2}\left( p\right) } . \label{lambda2}
\end{equation}

The condition of instability $d_{2}\left( p\right) <|g_{2}\left(
 p\right) |$ can be obeyed now even
for positive values of the scattering length (and hence of $U$), as $d_{2}$
 contains a
negative term $-\epsilon \left( p_{0}\right) $ [see Eq.\ (\ref{d2})].
 In the case of a contact
potential (\ref{Ua}) an instability takes place when
\begin{equation}
|\epsilon \left( p\right) -\epsilon \left( {\bf p}_{0}\right)
+{1\over 2}U\left( n_{+}+n_{-}\right) |<2U\sqrt{n_{+}n{ } _{-}} ,
 \label{inst2}
\end{equation}
or when
\begin{equation}
p_{\_}<p<p_{+}{ } _{,} \label{pinst}
\end{equation}
where
\begin{equation}
p_{\pm }=\left\lbrack p^{2}_{0}-m U \left( n_{+}+n_{-}\right)  \pm 4m
 U\sqrt{n_{+}n{ } _{-}}\right\rbrack ^{1/2} .
\end{equation}
Whenever condition (\ref{highp}) is obeyed, the momenta $p_{\pm }$
 are given approximately by
\begin{equation}
p_{\pm }\approx p_{0}-{mU\over p{ } _{0}} {n_{+}+n{ } _{-}\over 2}\pm
 {2mU\over p{ } _{0}}\sqrt{n_{+}n{ } _{-}} .
\end{equation}
The instability interval can include the region $p>p_{0}$. It may
 appear
that this fact can lead to the absurd conclusion that a ground-state
condensate ($p_{0}=0$) with a positive scattering length is unstable.
However, the analysis above is valid only when the condition
(\ref{highp}) is obeyed. Otherwise, the terms neglected in Eqs.\
(\ref{csip2}) and (\ref{phi2}) have to be taken into account, leading
 to
stability of the condensate at $p_{0}=0$.

As  in the case of a ground-state condensate, the scattered atoms
with momenta ${\bf p}$ and $-{\bf p}$ are correlated,
 forming optimally
entangled and optimally squeezed states [see discussion following Eq.\
(\ref{cor})].

\subsection{Elastic-scattering loss in collisions of two matter
waves} \label{Loss}

Consider now the elastic-scattering loss of atoms from the two
colliding
counterpropagating waves bounded only by the normalization box.
Substitution of Eqs.\ (\ref{den}) and (\ref{cor}) (with the
 replacement
of $d$, $g$, and $\lambda $ by $d_{2}$, $g_{2}$,  and $\lambda _{2}$,
 respectively) into Eq.\ (\ref{phi2})
leads to the following equations for the two condensate densities,
\begin{equation}
\dot{n}_{\pm }=-{m p_{0}U^{2}n_{+}n_-\over \pi { } ^{2}}\left(
 I_{\text{unst}}+I_{+}+I_{-}\right)  , \label{npm}
\end{equation}
where
\begin{eqnarray}
&&I_{\text{unst}}={1\over m p_0}\int\limits^{p{ } _{+}}_{p{ } _{
-}}p^{2}{\sinh\left( 2\lambda _{2}\left( p\right) t\right) \over
 \lambda _{2}\left( p\right) }d p \nonumber
\\
&&I_{-}={1\over m p_0}\int\limits^{p{ } _{-}}_{0}p^{2}{\sin\left( 2
|\lambda _{2}\left( p\right) |t\right) \over |\lambda _{2}\left(
 p\right) |}d p \label{Iunstpm}
\\
&&I_{+}={1\over m p_0}\int\limits^{\infty }_{p{ } _{
+}}p^{2}{\sin\left( 2|\lambda _{2}\left( p\right) |t\right) \over
|\lambda _{2}\left( p\right) |}d p \nonumber
\end{eqnarray}
represent the contributions of the unstable ($I_{\text{unst}}$) and stable
($I_{+}+I_{-}$) modes, respectively.

The substitution of
\[
p=\left\lbrack p^{2}_{0}-m U \left( n_{+}+n_{-}\right)  +4m
 U\sqrt{n_{+}n{ } _{-}}\cos \theta \right\rbrack ^{1/2}
\]
allows us to express $I_{\text{unst}}$  in terms of the modified Struve
 function
$L_{\nu }\left( z\right) $ (see Ref.\ \cite{BE53}), as
\begin{equation}
I_{\text{unst}}=\int\limits^{\pi }_{0}d\theta {p\left( \theta \right)
 \over p{ } _{0}}\sinh\left( \tau  \sin \theta \right) \approx \pi
 L_{0}\left( \tau \right)  , \label{Iinst2}
\end{equation}
where $\tau =2t/t_{\text{NL}}$,  and here $t_{\text{NL}}=\left( 2U\sqrt{n_{+}n{ } _{
-}}\right) ^{-1}$. Approximate expressions
for $I_{\text{unst}}$  at small and large values of $\tau $ have the
 respective forms
\begin{equation}
I_{\text{unst}}\mathrel{ \mathop \approx_{ \tau \rightarrow0}}2\tau
 ,\qquad I_{\text{unst}}\mathrel{ \mathop \approx_{ \tau \rightarrow
 \infty}} \sqrt{{\pi \over 2\tau }} e^{\tau }- {2\over \tau } .
 \label{Iinsta2}
\end{equation}
Given the condition $t p^{2}_{0}/m\ll 1$, the substitution of
 $|\lambda _{2}\left( p\right) |\approx p^{2}/\left( 2m\right) $
produces the estimate
\begin{equation}
I_{+}\approx {1\over p{ } _{0}}\sqrt{{\pi m\over 2 t}} .
 \label{Imnonst}
\end{equation}
The integral $I_{-}$  can be estimated under the same condition
by retaining only the first term in
the Taylor series for $\sin\left( 2|\lambda _{2}\left( p\right)
|t\right) $, giving $I_{-}\approx 2p^{2}_{0}t/\left( 3m\right) $.

Given the opposite condition,  $t p^{2}_{0}/m\gg 1$, the contributions
of the stable
 modes $I_{\pm }$
can be evaluated by substituting
\[
p=\left\lbrack p^{2}_{0}
-m U \left( n_{+}+n_{-}\right)  \pm 4m
U\sqrt{n_{+}n{ } _{-}}\cosh \theta \right\rbrack ^{1/2},
\]
resulting in
\begin{equation}
I_{\pm }=\int\limits^{\theta { } _{\pm }}_{0}d\theta {p\left( \theta
 \right) \over p{ } _{0}}\sin\left( \tau  \sinh \theta \right)  ,
\end{equation}
where $\theta _{+}=\infty $  and $p\left( \theta _{-}\right) =0$
(with $\theta _{-}\approx \ln\left\lbrack p^{2}_{0}/\left( 2mU\sqrt{n_{
+}n{ } _{-}}\right) \right\rbrack $). The total
contribution of the stable modes, given by
\begin{equation}
I_{+}+I_{-}\approx \pi \left( I_{0}\left( \tau \right) -L_{0}\left(
 \tau \right) \right)  , \label{Ipm}
\end{equation}
where $I_{0}\left( z\right) $ is the modified Bessel function (see
 Ref.\
\cite{BE53}), can be approximately expressed as \begin{mathletters}
\begin{eqnarray}
&&I_{+}+I_{-}\approx \pi -2\tau ,\qquad m/p^{2}_{0}\ll t\ll t_{\text{NL}}
 \label{Ipmst}
\\
&&I_{+}+I_{-}\approx 2/\tau ,\qquad t\gg t_{\text{NL}} . \label{Ipmlt}
\end{eqnarray}
\end{mathletters} Combining Eqs.\ (\ref{npm}), (\ref{Iinst2}),
 and (\ref{Ipm}),  one obtains the total rate of the
condensate loss
\begin{equation}
\dot{n}_{\pm }\approx -{m p_{0}U^{2}n_{+}n_-\over \pi }
 I_{0}\left( \tau \right) . \label{npmdota}
\end{equation}

Unlike the case of ground-state condensate, characterized by a single
time-scale $t_{\text{NL}}$, an additional time-scale, $m/p^{2}_{0}$,
appears in the present case, leading to appearance of a third
regime of the condensate loss.
At short times, $t\ll t_{\text{NL}}$,  the loss is dominated by the
occupation of the
stable modes due to their higher statistical weight.

At the regime of non-stationarity $t p^{2}_{0}/m\ll 1$,
the wavefunction of the
 scattered particles
does not have enough time to become stationary, and the estimate
(\ref{Imnonst}) gives the loss rate
\begin{equation}
\dot{n}_{\pm }\approx -{m U^{2}n_{+}n_-\over \pi }
 \sqrt{{m\over 2 \pi t}} . \label{npmdotn}
\end{equation}
The singularity at
$t=0$ has the same nature as in the case of the ground-state condensate
with a negative scattering length [see discussion following Eq.\
(\ref{Deltan})].

At the regime of stationarity, $m/p^{2}_{0}\ll t\ll t_{\text{NL}}$,  the
loss rate becomes time independent, and given by
\begin{equation}
\dot{n}_{\pm }\approx -16\pi a^{2}_{a}{p{ } _{0}\over m}n_{+}n_{-}
 ,\label{ndotstat}
\end{equation}
in full agreement with  theory \cite{BTBJ00}.

At the regime of Bose-enhancement, $t\gg t_{\text{NL}}$,  the loss is
principally determined by the
occupation of unstable modes which increase exponentially due to the
 effect
of Bose enhancement. The loss rate can then be approximately expressed as
\begin{equation}
\dot{n}_{\pm }\approx -{\left( 2a_{a}\right) ^{3/2}p_{0}\left( n_{
+}n_{-}\right) { } ^{3/4}\over \sqrt{mt}}\exp\left( 2t/t_{\text{NL}}\right)
 . \label{ndotenh}
\end{equation}

\begin{figure}
\psfig{figure=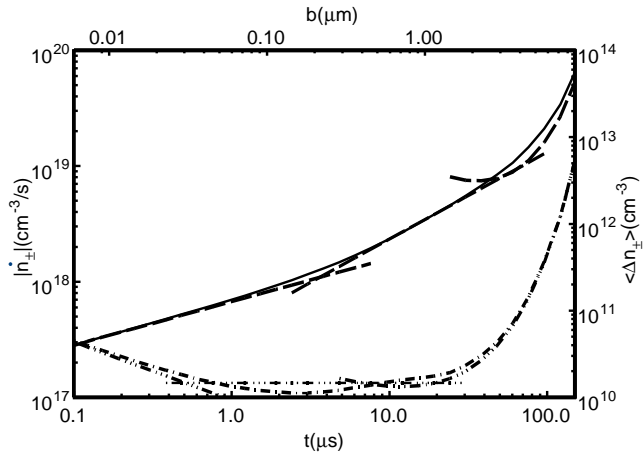,width=3.375in}

\caption{Time-dependence of the condensate loss rate $\dot{n}_{\pm}$
under conditions of the Na NIST  experiment \protect\cite{D99}
($n_{\pm}\approx 10^{14}$cm$^{-3}$). The results of numerical integration
in Eqs.\ (\protect\ref{Iunstpm})
(dot-dashed line) are compared to the approximate equations
(\protect\ref{npmdotn}), (\protect\ref{ndotstat}), and (\protect\ref{ndotenh})
(dot-dot-dot-dashed line, left, central,
and right segments, respectivelly).
The solid line represents
the average density loss $\Delta n_0$ for the collision of rectangular
wavepackets calculated by using of numerical integration in
Eqs.\ (\protect\ref{Iunstpm}) and (\protect\ref{Dnpma}),
in comparison with the approximate expressions (\protect\ref{Dnpmn}),
(\protect\ref{Dnpms}), and (\protect\ref{Dnpmbe}) (dashed line, left,
central, and right segments, respectivelly). \label{Fc}}

\end{figure}

The approximate expressions (\ref{npmdotn}), (\ref{ndotstat}),
and (\ref{ndotenh}) are
compared in Fig.\ \ref{Fc} to results of numerical integration in
Eqs.\ (\ref{Iunstpm}).

The contribution of stable modes decreases at long times as $t$
increases [see Eq.\
(\ref{Ipmlt})]. In order to explain the decrease, let us consider the
asymptote of $I_{+}+I_{-}$  using, in the spirit of scattering theory
 \cite{GW},
the substitution
\begin{eqnarray}
\lim_{t\rightarrow \infty }&&\sin\left( 2|\lambda _{2}\left( p\right)
 |t\right) /|\lambda _{2}\left( p\right) |=\pi \delta \left( \lambda
 _{2}\left( p\right) \right)  \nonumber
\\
&&=\pi \left\lbrack {\delta \left( p-p_{+}\right) \over \left|{d\over
 d p}|\lambda _{2}\left( p_{+}\right) |\right|}+{\delta \left( p-p_{
-}\right) \over \left|{d\over d p}|\lambda _{2}\left( p_{-}\right)
|\right|} \right\rbrack .
\end{eqnarray}
This leads to
\begin{equation}
I_{+}+I_{-}\sim {\pi \over 2mp{ } _{0}}\left\lbrack {p{ } ^{2}_{
+}\over \left|{d\over d p}|\lambda _{2}\left( p_{+}\right) |\right|}
+{p{ } ^{2}_{-}\over \left|{d\over d p}|\lambda _{2}\left( p_{
-}\right) |\right|}\right\rbrack  .
\end{equation}
In the absence of many-body effects ($U=0$) , one obtains $|\lambda _{2}\left(
 p\right) |=|p^{2}-p^{2}_{0}|/\left( 2m\right) $,
expressing the regular dispersion law for free particles. Also, $p_{\pm
 }=p_{0}$,
$\left|{d\over d p}|\lambda _{2}\left( p_{\pm }\right) |\right|=p_{0}
/m$, and hence $I_{+}+I_{-}=\pi $, which also hold for small $t$ [see
Eq.\ (\ref{Ipmst})]. Therefore the scattering process can be
 described
by a stationary scattering theory at any time $t\gg m/p^{2}_{0}$
(see Ref.\
\cite{GW}). Contrarily, in the case of a finite $U$,  many-body effects
 modify
the dispersion law, so that $\left|{d\over d p}|\lambda _{2}\left(
 p_{\pm }\right) |\right|\rightarrow \infty $, and hence
$I_{+}+I_{-}\rightarrow 0$.

\subsection{Collisions of rectangular wavepackets} \label{Rect}

The discussion above is related to the scattering of two condensate
waves bounded only by a normalization box. A more realistic yet
simple model is that of a half-collision of rectangular wavepackets. Consider
 a
homogeneous condensate occupying a region of  length $b$ along the
 $x$
axis (leaving the results  independent of size and  shape in the
$yz$-plane). At the moment $t=0$ a Bragg scattering pulse is applied,
creating a
wavepacket of density $n_{+}$  and momentum $2{\bf p}_{0}$
 directed along the $x$
axis, while the rest of the atoms,  with density $n_{-}$,  remains
in the ${\bf p}=0$ wavepacket.
Losses take place only at the overlapping parts of the
 two wavepackets.
The overlapping time $t_{\text{over}}$  varies from 0 to
 $t_{\text{col}}$,  between the leading
and trailing edges of the moving wavepacket, where
\begin{equation}
t_{\text{col}}={m b\over 2p{ } _{0}}
\end{equation}
is the total duration of the collision of the two wavepackets. The
average density loss  can be expressed as
\begin{equation}
\langle \Delta n_{\pm }\rangle =-{1\over t{ }
 _{\text{col}}}\int\limits^{t{ }
 _{\text{col}}}_{0}dt_{\text{over}}\int\limits^{t{ }
 _{\text{over}}}_{0}dt \dot{n}_{\pm }\left( t\right)  , \label{Dnpma}
\end{equation}
where $\dot{n}_{\pm }\left( t\right) $ is defined by Eq.\
(\ref{npmdota}).

At the regime of non-stationarity, $t_{\text{col}}p^{2}_{0}/m\ll 1$, or
 $b p_{0}\ll 1$, the loss
rate is given by Eq.\ (\ref{npmdotn}), and
\begin{equation}
\langle \Delta n_{\pm }\rangle \approx {32\over 3} \sqrt{{\pi b\over
 p{ } _{0}}}a^{2}_{a}n_{+}n_{-} . \label{Dnpmn}
\end{equation}
At the regime of stationarity,
$m/p^{2}_{0}\ll t_{\text{col}}\ll t_{\text{NL}}$,  the loss rate is
 determined by Eq.\ (\ref{ndotstat}), and the loss has a form
\begin{equation}
\langle \Delta n_{\pm }\rangle \approx 4\pi a^{2}_{a}b n_{+}n_{-} .
\label{Dnpms}
\end{equation}
The same expression for the loss can also be obtained by using the
method of complex scattering length \cite{BTBJ00}.

At the regime of Bose-enhancement,
$t_{\text{col}}\gg t_{\text{NL}}$,  the loss
 rate Eq.\ (\ref{ndotenh}) gives the average density loss
\begin{equation}
\langle \Delta n_{\pm }\rangle \approx {p{ } ^{5/2}_{0}\over 32\pi
 ^{2}a^{1/2}_{a}b^{3/2}\left( n_{+}n_{-}\right) { } ^{1/4}}\exp\left(
 2t_{\text{col}}/t_{\text{NL}}\right) \label{Dnpmbe}
\end{equation}
The approximate expressions (\ref{Dnpmn}), (\ref{Dnpms}),
and (\ref{Dnpmbe}) are
compared in Fig.\ \ref{Fc} to results of numerical integration in
Eqs.\ (\ref{Iunstpm}) and (\ref{Dnpma}).

The parametric approximation and, hence all the analysis above,
are applicable only as long as $\langle \Delta n_{\pm }\rangle \ll
 \sqrt{n_{+}n{ } _{-}}$. Using Eq.\ (\ref{Dnpmbe}), this
condition can be rewritten in the form of a limit on the duration of
collision consistent with the approximation:
\begin{equation}
t_{\text{col}}<{1\over 2}t_{\text{NL}}\ln{32\pi ^{2}a^{1/2}_{a}b^{3/2}\left(
 n_{+}n_{-}\right) { } ^{3/4}\over p{ } ^{5/2}_{0}} .
\end{equation}
For example, in the NIST Na experiment \cite{D99} (where $n_{+}\approx n_{
-}\approx 10^{14}$cm$^{-3}$,
$a_{a}\approx 3$nm, $b\approx 20\mu $m, and $p_{0}\approx 10^{5}$cm$^{
-1}$; see Ref.\ \cite{TBJ00}), the parametric
approximation is applicable while $t_{\text{col}}<3t_{\text{NL}}$, which
 allows for values of the
exponential Bose-enhancement factor in Eq.\ (\ref{Dnpmbe}) of more
 than
a 100. This validity range substantially exceeds the range of
applicability of the complex scattering-length method,
 $t_{\text{col}}\ll t_{\text{NL}}$.
However, in the four-wave-mixing NIST experiment \cite{D99},
$t_{\text{col}}\approx 7t_{\text{NL}}$, which requires taking
into account a considerable depletion of the condensates during the
collision. This
task may be performed by combining  the present approach with that of the
slowly-varying-envelope approximation \cite{BTBJ00,TBJ00}.

\subsection{Collisions of Gaussian wavepackets} \label{Gauss}

The consideration of wavepackets of a more realistic shape requires
the
solution of Eq.\ (\ref{csip2}) for time-dependent $d_{2}$  and
 $g_{2}$. This
solution can be obtained by using a perturbation theory, which is
applicable as long as $t\ll t_{\text{NL}}$,  and does not describe the effects of Bose
enhancement. The evaluation is made simpler in the coordinate
representation, in which case the linearized equations of motion for the
field operators of the fluctuations  $\hat{\tilde{\xi }}\left( {\bf
 r},t\right) $ have the form
\begin{eqnarray}
&&i \dot{\hat{\tilde{\xi }}}\left( {\bf r},t\right) ={\hat{p}{ }
 ^{2}\over 2m}\hat{\tilde{\xi }}\left( {\bf r},t\right)
+ N\int d^{3}r^\prime \tilde{U}\left( {\bf r}-{\bf
 r}^\prime \right)  \Bigl\lbrack |\varphi \left( {\bf r}^\prime
 ,t\right) |^{2}\hat{\tilde{\xi }}\left( {\bf r},t\right)  \nonumber
\\
&&+\varphi ^{*}\left( {\bf r}^\prime ,t\right) \varphi \left( {\bf
 r},t\right) \hat{\tilde{\xi }}\left( {\bf r}^\prime ,t\right)
+\varphi \left( {\bf r},t\right) \varphi \left( {\bf r}^\prime
 ,t\right) \hat{\tilde{\xi }}^{\dag }\left( {\bf r}^\prime ,t\right)
 \Bigr\rbrack  , \label{csir}
\end{eqnarray}
where $\varphi \left( {\bf r},t\right) $ is a unit-normalized
 wavefunction describing
particles in the colliding wavepackets, $N$ is the total number of
particles in the wavepackets, and $\tilde{U}\left( {\bf r}\right) $
is the interatomic interaction potential.

The first-order perturbative solution of Eq.\ (\ref{csir}) can be
expressed in the form
\begin{eqnarray}
\hat{\tilde{\xi }}_{1}\left( {\bf r},t\right) =\int d^{3}r^\prime
 \int\limits^{t}_{0}dt^\prime \bigl\lbrack A\left( {\bf r},t;{\bf
 r}^\prime ,t^\prime \right) \hat{\tilde{\xi }}_{0}\left( {\bf
 r}^\prime ,t^\prime \right)  \nonumber
\\
+B\left( {\bf r},t;{\bf r}^\prime ,t^\prime \right) \hat{\tilde{\xi
 }}^{\dag }_{0}\left( {\bf r}^\prime ,t^\prime \right) \bigr\rbrack .
\label{xi1}
\end{eqnarray}
Here the zero-order solution
\begin{equation}
\hat{\tilde{\xi }}_{0}\left( {\bf r},t\right) =\sum\limits^{}_{{\bf
 p}}\hat{\xi }\left( {\bf p},0\right) \chi _{{\bf p}}\left( {\bf
 r},t\right)
\end{equation}
is expressed in terms of free (plane) waves
\begin{equation}
\chi _{{\bf p}}\left( {\bf r},t\right) ={\cal V}^{-1/2}\exp\left( i{\bf
 p}{\bf r}-i{{\bf p}{ } ^{2}\over 2m}t\right)
\end{equation}
and the annihilation operators $\hat{\xi }\left( {\bf p},t\right) $
 for the fluctuation in the
momentum representation.
As the following equations turn out to be independent of the first
term in Eq.\ (\ref{xi1}), including the function
$A\left( {\bf r},t;{\bf r}^\prime ,t^\prime \right) $ , it is not
necessary to specify that function. The second term includes the
function
\begin{eqnarray}
B\left( {\bf r},t;{\bf r}^\prime ,t^\prime \right) =-i N\int
 d^{3}r^{\prime\prime}\sum\limits^{}_{{\bf p}}\chi _{{\bf p}}\left(
 {\bf r},t\right) \chi ^{*}_{{\bf p}}\left( {\bf
 r}^{\prime\prime},t^\prime \right)  \nonumber
\\
\times \tilde{U}\left( {\bf r}^{\prime\prime}-{\bf r}^\prime \right)
 \varphi \left( {\bf r}^{\prime\prime},t^\prime \right) \varphi \left
( {\bf r}^\prime ,t^\prime \right)  .
\end{eqnarray}

Considering scattering into the vacuum of fluctuations, such that
$\langle \hat{\tilde{\xi }}^{\dag }_{0}\left( {\bf r},t\right)
 \hat{\tilde{\xi }}_{0}\left( {\bf r}^\prime ,t^\prime \right)
 \rangle =\langle \hat{\tilde{\xi }}_{0}\left( {\bf r},t\right)
\hat{\tilde{\xi }}_{0}\left( {\bf r}^\prime ,t^\prime \right) \rangle
 =0$, and therefore
\begin{equation}
\langle \hat{\tilde{\xi }}_{0}\left( {\bf r},t\right)
 \hat{\tilde{\xi }}^{\dag }_{0}\left( {\bf r}^\prime ,t^\prime \right
) \rangle =\sum\limits^{}_{{\bf p}}\chi _{{\bf p}}\left( {\bf
 r},t\right) \chi ^{*}_{{\bf p}}\left( {\bf r}^\prime ,t^\prime
 \right),
\end{equation}
 one can express the number of the lost atoms in the form
\begin{eqnarray}
\Delta N=&&-\int d^{3}r\langle \hat{\tilde{\xi }}^{\dag }_{1}\left(
 {\bf r},\infty \right) \hat{\tilde{\xi }}_{1}\left( {\bf r},\infty
 \right) \rangle  \nonumber
\\
=&&-\int d^{3}r \int d^{3}r_{1}\int d^{3}r_{2}\int\limits^{\infty
 }_{0}dt_{1}\int\limits^{\infty }_{0}dt_{2}B^{*}\left( {\bf r},\infty;{\bf
 r}_{1},t_{1}\right)  \nonumber
\\
&&\times B\left( {\bf r},\infty;{\bf r}_{2},t_{2}\right)
 \sum\limits^{}_{{\bf p}}\chi _{{\bf p}}\left( {\bf
 r}_{1},t_{1}\right) \chi ^{*}_{{\bf p}}\left( {\bf
 r}_{2},t_{2}\right)  \nonumber
\\
=&&2N^{2}\text{Im}T^{\left( 2\right) }_{00}
\end{eqnarray}
where
\begin{eqnarray}
T^{\left( 2\right) }_{00}&&=\int\limits^{\infty
 }_{0}dt^{\prime\prime}\int\limits^{\infty }_{0}dt^\prime \int
 d^{3}r_{1}^\prime d^{3}r_{2}^\prime
 d^{3}r_{1}^{\prime\prime}d^{3}r_{2}^{\prime\prime}
\varphi ^{*}\left( {\bf r}_{1}^{\prime\prime},t^{\prime\prime}\right)
\nonumber
\\
&&\times  \varphi ^{*}\left(
 {\bf r}_{2}^{\prime\prime},t^{\prime\prime}\right) \tilde{U}\left(
 {\bf r}_{1}^{\prime\prime}-{\bf r}_{2}^{\prime\prime}\right) G\left(
 {\bf r}_{1}^{\prime\prime},{\bf
 r}_{2}^{\prime\prime},t^{\prime\prime};{\bf r}_{1}^\prime ,{\bf
 r}_{2}^\prime ,t^\prime \right)  \nonumber
\\
&&\times \tilde{U}\left( {\bf r}_{1}^\prime -{\bf r}_{2}^\prime
 \right) \varphi \left( {\bf r}_{1}^\prime ,t^\prime \right) \varphi
 \left( {\bf r}_{2}^\prime ,t^\prime \right)  . \label{DeltaN}
\end{eqnarray}
Here
\begin{eqnarray}
G&&\left( {\bf r}_{1}^{\prime\prime},{\bf
 r}_{2}^{\prime\prime},t^{\prime\prime};{\bf r}_{1}^\prime ,{\bf
 r}_{2}^\prime ,t^\prime \right) =-i\sum\limits^{}_{{\bf p}_{1},{\bf
 p}{ } _{2}}\chi _{{\bf p}_1}\left( {\bf
 r}_{1}^{\prime\prime},t^{\prime\prime}\right) \chi _{{\bf p}_2}\left
( {\bf r}_{2}^{\prime\prime},t^{\prime\prime}\right)  \nonumber
\\
&&\times \chi ^{*}_{{\bf p}_1}\left( {\bf r}_{1}^\prime ,t^\prime
 \right) \chi ^{*}_{{\bf p}_2}\left( {\bf r}_{2}^\prime ,t^\prime
 \right) \vartheta \left( t^{\prime\prime}-t^\prime \right)
\end{eqnarray}
and $\vartheta \left( t\right) $ is the Heavyside step function.
As one can see $G( {\bf r}_{1}^{\prime\prime},{\bf r}_{2}^{\prime\prime},
t^{\prime\prime};{\bf r}_{1}^\prime ,{\bf r}_{2}^\prime ,t^\prime )$
is a time-dependent Green function for two free particles.
Therefore $T^{\left( 2\right) }_{00}$  is nothing else
but the second-order
term for the diagonal matrix element of the transition
amplitude, obtained by time-dependent perturbation theory for a
collision of two particles, each in the state $\varphi \left( {\bf
 r},t\right) $. If the size of the
wavepacket obeys  $b\gg 1/p_{0}$, the result can be reduced to a
form given by the
stationary scattering theory (see Ref.\ \cite{GW}), which is
proportional to the imaginary part of the elastic scattering length,
in agreement with Ref.\ \cite{BTBJ00}. Consider two colliding Gaussian
wavepackets described by the wavefunction
\begin{eqnarray}
\varphi \left( {\bf r},t\right) =&&\left\lbrack F_{+}\left( {\bf r}
-{{\bf p}{ } _{0}\over m}t\right) e^{i{\bf p}_0{\bf r}}+F_{-}\left(
 {\bf r}{\bf +}{{\bf p}{ } _{0}\over m}t\right) e^{-i{\bf p}_0{\bf
 r}}\right\rbrack  \nonumber
\\
&&\times \exp\left( -{p{ } ^{2}_{0}\over 2m}t\right)  , \label{phiG}
\end{eqnarray}
where
\begin{eqnarray}
F_{\pm }\left( {\bf r}\right) =&&\pi ^{-3/4}\sqrt{N_{\pm }/N} \left(
 b_{x}b_{y}b_{z}\right) ^{-1/2} \nonumber
\\
&&\times \exp\left( -{x{ } ^{2}\over 2b{ } ^{2}_{x}}-{y{ } ^{2}\over
 2b{ } ^{2}_{y}}-{z{ } ^{2}\over 2b{ } ^{2}_{z}}\right)
\end{eqnarray}
and $N_{\pm }$  is the number of particles in each wavepacket.
Substituting Eq.\ (\ref{phiG}) into Eq.\ (\ref{DeltaN}) one can obtain
\begin{equation}
\Delta N=4{N_{+}N{ } _{-}\over b_{x}b_{y}b{ } _{z}}{p{ } _{0}\over
 m}a^{2}_{a}t_{\text{col}} , \label{DeltaNst}
\end{equation}
where
\begin{equation}
t_{\text{col}}=m\left( {p{ } ^{2}_{0x}\over b{ } ^{2}_{x}}+{p{ }
 ^{2}_{0y}\over b{ } ^{2}_{y}}+{p{ } ^{2}_{0z}\over b{ }
 ^{2}_{z}}\right) ^{-1/2}
\end{equation}
is the collision duration. Equation (\ref{DeltaNst})  coincide
with the expression which can be obtained by using of the complex
scattering length method \cite{BTBJ00} and neglecting the depletion of
the wavepackets.

The recent preprint \cite{BTR01}, posted during preparation of the present
paper, treats the loss of atoms from colliding BEC wavepackets by
using an approach similar to the  one presented here, with solutions
of the form
of Eq.\ (\ref{solap}) published previously in Ref.\ \cite{VYA01}.
However, unjustified neglections of substantial terms in the
 Hamiltonian
lead to an expression for $g_{2}$  which is smaller by a factor of
 four than the one given here by
Eq.\ (\ref{g2}). This neglection should therefore result in an
underestimation of the condensate loss.

The above-mentioned work, Ref. \cite{BTR01}, considers collision of
Gaussian wavepackets
to first-order perturbation theory only, which do not describe the
effects of Bose enhancement. Finite analytical expressions of the
kind presented here were not
derived there, and the results of numerical calculations disagree
 with
results of the complex scattering length method \cite{BTBJ00} even
after the correcting for the missing factor four in $g_{2}$
are taken into account.
In contrast, the analytical expression Eq.\ (\ref{DeltaNst}) of the
present paper, based on the same approximations,
is in an agreement with the complex scattering length method.

\section{Conclusions}

Fluctuations in BEC are treated by using a parametric
approximation for the exact quantum equations of motion for the field
operators. Depending on the excitation energy, the fluctuations
are divided into stable modes, which can be represented as Bogolubov
quasiparticles (field oscillators), and unstable modes, which can be
represented as inverted oscillators and demonstrate a rapid growth.
Initially the dynamics of unstable modes is a quantum process of
spontaneous decay into the vacuum of the fluctuations with a
super-exponential rate.

This approach is applied to two physical problems. The first one
involves the dynamics of a BEC after a fast change of the scattering length
from a non-negative  to a negative one. In this case , the
experimentally observed \cite{D01} delay of the condensate collapse
is described.
The second problem involves the elastic scattering losses on collision of two
BEC wavepackets. At short collision times [see Eq.\
(\ref{ndotstat})] the losses obtained here are in agreement with those
of the
method of complex scattering length   \cite{BTBJ00} , with specific
reference to the collisions of
non-confined waves, of rectangular and of  Gaussian wavepackets. At longer
collision times [see Eq.\ (\ref{ndotenh})]  quantum effects lead to
Bose-enhancement of the elastic scattering losses. The atoms lost from
the condensates in both cases form correlated pairs in optimally entangled
and optimally squeezed states.

\acknowledgments

The author is very grateful to P.\ S.\ Julienne and K.\ Burnett for
helpful discussions. Special appreciation is expressed to
A.\ Ben-Reuven
for invaluable discussions and the help in preparation of the present
paper. This work was partially supported by the National Science
 Foundation
through a grant for the Institute for Theoretical Atomic and Molecular
Physics at Harvard University and the Smithsonian Astrophysical
Observatory.

\newpage

\end{document}